# Ensemble Models for Detecting Wikidata Vandalism with Stacking

Team Honeyberry Vandalism Detector at WSDM Cup 2017

Tomoya Yamazaki, Mei Sasaki, Naoya Murakami, Takuya Makabe, Hiroki Iwasawa
Yahoo Japan Corporation
{tomoyama, mesasaki, naomurak, tmakabe, hiwasawa}@yahoo-corp.jp

## ABSTRACT

The WSDM Cup 2017 is a binary classification task for classifying Wikidata revisions into vandalism and non-vandalism. This paper describes our method using some machine learning techniques such as under-sampling, feature selection, stacking and ensembles of models. We confirm the validity of each technique by calculating AUC-ROC of models using such techniques and not using them. Additionally, we analyze the results and gain useful insights into improving models for the vandalism detection task. The AUC-ROC of our final submission after the deadline resulted in 0.94412.

## 1. INTRODUCTION

With the growing number of edits on public knowledge bases, the number of ill-intentional/erroneous edits has increased as well. Thus, detecting such vandalizing or damaging edits is essential in the further development of knowledge bases. The WSDM Cup 2017 [1, 4] challenge requires participants to measure the probability that each Wikidata revision is damaging or edited by vandalism.

We develop vandalism detection models in the following four steps: (1) preprocessing data, (2) extracting features, (3) feature engineering, and (4) training models. Some of the difficulties we faced in this competition were inherent in the WSDM Cup 2017 task: near real-time classification of revisions, the size of the data set, and the relative infrequency of vandalizing revisions compared to non-vandalizing revisions, namely heavily imbalanced data. Other difficulties had a more technical aspect and were related to software submission using TIRA [6].

To solve such problems and improve our models, we adopted a python client to connect to the WSDM server that provides evaluation data and a python code for training and prediction. The machine learning techniques we used include: **under-sampling** for handing huge and imbalanced data, **feature selection** for improving our results and reducing training time, **stacking** and **ensemble** technique frequently used in machine learning competitions for improving models by avoiding over-fitting.

This paper is organized as follows: we describe our method that consists of the aforementioned four steps in more detail in Section 2, we show results of our experiments on a given validation data in Section 3, we explain analysis results in Section 4, and we conclude the paper and explain our future work in Section 5.

## 2. METHODS

All experiments were conducted on a Mac OS X with Intel Core i7 (2.2 GHz) and 16 GB of main memory. Our method was implemented in Python.

### 2.1 Preprocessing Data

#### 2.1.1 Data Description

The WSDM Cup 2017 distributes Wikidata revision data from Oct 2012 to Apr 2016[1]. For validation, we use the data set from Oct 2012 to Feb 2016 as training data and the data set of Mar and Apr 2016 as validation data. The available revision data is very huge and heavily imbalanced; for example, the number of vandalizing data (**positive**), and not vandalizing data (**negative**) of the validation data set are 10,784 and 7,213,861, respectively.

In our environment we cannot handle the entire data provided, thus we separate the training data into positive and negative examples and use all positive examples and negative examples only in the data set of Jan and Feb 2016 for validation or Mar and Apr 2016 for the final submission. Since the revision data sets are XML format and the file that contains the user information is separated from the revision data, we had to parse the given data set and refer to another file. Thus, we extract revision IDs, comments, binary values whether the revision contains contributors tag and user information and merge them into a single file to avoid the time-consuming work, extracting features from the raw data set. The merged file contains lines such as `"308612969 /* wbsetclaim-create:2||1 */ [[Property:P800]]: [[Q5974487]] 0,GB,EU,GMT,EN,LEEDS,WEST YORKSHIRE,"`. All following experiments were conducted by using the training data set.

#### 2.1.2 Under-Sampling

Since learning algorithms that do not adopt to imbalanced data tend to be overwhelmed by the major class and ignore the minor class [2], we employ an under-sampling technique for negative data and adjust the ratio of positive and negative data. Our simple and fast sampling method is composed of two steps; (1) random sampling with a sampling fraction from negative samples; (2) removing duplicate contents of all data. To confirm the effectiveness of the both of the steps, we measure the AUC-ROC with default random forest and compare the score between the model with each of the steps and the model without each of them.

Figure 1 depicts the comparison of AUC-ROC among different sampling fraction and shows using $\frac{1}{50}$ sampling data set of the latest negative data set results in the best scores. Scores shown in Figure 1 use the data set that removed duplicates. We also confirm the effectiveness of removing duplicates with $\frac{1}{50}$ sampling data set, because the AUC-ROC increases from 0.94678 to 0.95124 by removing duplicate contents. Hereinafter, in our experiments, we use the latest $\frac{1}{50}$ sampling negative samples and all positive examples without duplicate contents as the training data set.

---

[1] http://www.wsdm-cup-2017.org/vandalism-detection.html

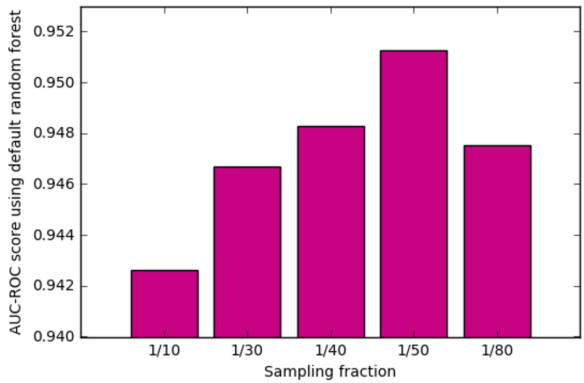

**Figure 1:** Comparison of AUC-ROC among the different sampling fraction with default random forest.

**Table 1:** Features proposed by Heindorf et al. [3] we adopted

| Feature category | Used features | |
|---|---|---|
| Content features | lowerCaseRatio<br>nonLatinRatio<br>alphanumericRatio<br>punctuationRatio<br>longestCharSeq<br>lowerCaseWordRatio<br>containsLangWord<br>upperCaseWordRatio | upperCaseRatio<br>latinRatio<br>digitRatio<br>whitespaceRatio<br>langWordRatio<br>containsURL<br>longestWord |
| Context features | userCountry<br>userCity<br>userRegion<br>isRegisteredUser<br>revisionLanguage<br>commentLength<br>revisionSubaction | userTimeZone<br>userCounty<br>userContinent<br>revisionTag<br>revisionAction<br>isLatinLanguage |

## 2.2 Feature Engineering

In this section, we give an explanation how to transform the sampling raw data to create features of the input of our models.

### 2.2.1 Extracting Features

We use a subset of features adopted in [3] and shown in Table 1. We treat context features other than `commentLength` as categorical variables, thus we convert these features to one-hot vectors by using the version 0.18.1 of scikit-learn[2], an open source Python library. In addition, we fill in missing values by zero values.

In addition to the features shown in Table 1, we extract other features shown in Table 2. The `containsHashTag` and `isSpecContriUser` are binary features that represent whether the comment includes hash tag (e.g. #autolist), and whether a special contribution user tag (e.g. [[Special:Contributions/abcd]]), a subset of "revisionAction" feature, is, respectively. We confirm the validity of features shown in Table 2 by measuring AUC-ROC using the default scikit-learn random forest with validation data set.

We further experimented other features: `isProperty` and `isPropertyQuestion` whether the comment contains property tags (e.g. [[Property:P641]]), and whether the comment contains question tags (e.g. [[Q41466]]), respectively, however the features do not contribute to improving the result.

[2]http://scikit-learn.org/

**Table 2:** Comparison of AUC-ROC among new defined features.

| Feature category | AUC-ROC |
|---|---|
| containsHashTag | 0.770 |
| isSpecContriUser | 0.520 |

### 2.2.2 Feature Selection

Since we convert label features to one-hot vectors, the number of features increase from 30 to 1,279 and the feature matrix is sparse, to be specific, our model trains and predicts by using 1,279 features. We use SelectFromModel module with the threshold used is default (e.g. 1e-5) and a gradient boosting tree package of a scikit-learn as a feature selection model and select 53 features when we set the random_state, one of the hyper parameters of the gradient boosting tree, as 0. The AUC-ROC of our models with feature selection increase from 0.93731 to 0.95124 using the same model and the data set as Section 2.1.2, furthermore, the models decrease the training time from about 3 hours to 10 minutes.

### 2.2.3 Stacking

Stacking (also called stacked generalization) is proposed by Wolpert [7] and is a popular method for improving results on data mining competitions. The training algorithm of 3-fold stacking is as follows:

1. Split the training data into 3 parts represented as X, Y and Z.
2. Fit a first-stage model on Y and Z, and calculate prediction score for X.
3. Fit the same first-stage model on X and calculate prediction score for Y and Z in the same manner as the previous step.
4. Fit a second-stage model on the union of prediction scores of X, Y and Z from the first-stage models.

For prediction, we calculate prediction scores for the test data using the fitted models. Some of the advantages of the stacking techniques are, for example, avoiding over-fitting due to K fold cross validation and interpreting non-linearity between features due to treating output scores as features.

We compare AUC-ROC among 5 models using stacking and without stacking and confirm the effectiveness shown in Table 3, the following section describes the details.

## 2.3 Training Models

Our final model is composed of 3-layer learning architecture as shown in Figure 2. Hereinafter, models tuned hyper parameters using grid search and not tuned models are referred to as optimized and default models, respectively. On the first-stage, we convert raw features described in Section 2.2.1 to 6 dimensional vectors by using 3-fold stacking technique and 6 different learning models of scikit-learn library as follows:

1. Default Multi-layer Perceptron Classifier (MLP)
2. Default Extra Trees Classifier
3. Extra Trees Classifier whose number of estimators is 200
4. Default Gradient Boosting Classifier
5. Gradient Boosting Classifier whose number of estimators is 200

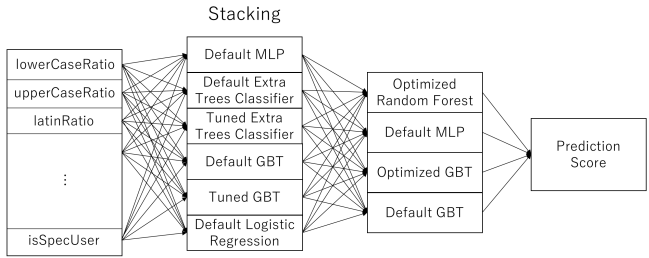

Figure 2: Overview of our final model architecture

Table 3: Comparison of AUC-ROC between models using stacking and single models without using stacking. RF, MLP, GBT and Ensemble means Random Forest, Multi-layer Perceptron, Gradient Boosting Trees and calculating the mean values of all models we use, respectively. Optimized models are tuned hyper parameters using grid search.

| Models | Stacking model | Single model |
|---|---|---|
| Optimized RF | 0.95898 | 0.95334 |
| Default MLP | 0.95778 | 0.95311 |
| Optimized GBT | 0.95774 | 0.95391 |
| Default GBT | 0.95564 | 0.95214 |
| Ensemble | 0.95920 | 0.95527 |

6. Default Logistic Regression

On the second-stage, we predict the final probabilities using the stacked features by using 4 different learning models as follows:

1. Optimized Random Forest Classifier whose number of estimators is 200 and max depth is 8
2. Default Multi-layer Perceptron Classifier (MLP)
3. Optimized Gradient Boosting Classifier whose number of estimators is 200 and max depth is 6
4. Default Gradient Boosting Classifier

On the last-stage, we calculate the mean value of the values of the second-stage.

Training the models and predicting the validation data set took 30 minutes in spite of our environments, standard personal computers. Note that training and predicting without stacking took ~10 minutes.

Table 3 shows the comparison of AUC-ROC between the final models on the second-stage and the effectiveness of our model compared to the models without using stacking.

We further experimented with default LightGBM package[3] proposed by Meng et al. [5] as the stacking and the final model, however the LightGBM does not improve the performance.

## 3. EVALUATION RESULTS

In this section, we show the results by evaluating the validation data set described in sectioin 2.1.1. We compare AUC-ROC among employing feature selection, stacking or ensemble with the validation data set, and show in Table 4. Since models using stacking without feature selections require much time, we have not conducted the experiments. Table 4 shows feature selection, stack-ing and ensemble techniques improve the performance. The model combined all the techniques outperform all results of the validation data sets.

[3]https://github.com/Microsoft/LightGBM

Table 4: Comparison of AUC-ROC among models using feature selection, stacking and ensemble techniques. Since the models not using feature selection but using stacking require much times, we do not conduct the experiments.

| Feature selection | Stacking | Ensemble | AUC-ROC |
|---|---|---|---|
| × | × | × | 0.95180 |
| × | × | ✓ | 0.95315 |
| ✓ | × | × | 0.95391 |
| ✓ | × | ✓ | 0.95527 |
| ✓ | ✓ | × | 0.95774 |
| ✓ | ✓ | ✓ | **0.95920** |

Unfortunately, we submitted our software included one critical bug, thus the final score was 0.90487. We convert label features to one-hot vectors using OneHotEncoder function of scikit-learn. The function accept not string but integer, thus we had to convert labels to unique integers using the hash function of a hashlib module. Clearly, there is no guarantee that hashed values of specific labels are the same in different environments. After the deadline, we fixed the bug by using original hash functions which can convert the string to unique integer in different environments and resubmitted our software via TIRA and which resulted in an unofficial score of 0.94412.

## 4. ANALYSIS

We analyze incorrect predicted revision data of our result.

### 4.1 Duplicated contents

Figure 3 depicts the number of incorrect predicted revisions. The upper panel of Figure 3 indicates almost all results can be predicted properly. The lower panel shows to be biased statics, which imply the number of the similar feature vectors that our model cannot predict properly is large. For instance, false positive results contain the 1,217 same revisions whose comments and user information are "/* wbeditentity-update:0| */" and "US,NA,EST,NJ,WOODBRIDGE,MIDDLESEX,", respectively, and false negative results contain the 118 same revisions whose comments and user information are "/* wbcreateclaim-create:1| */ [[Property:P106]]: [[Q47064]], #autolist2" and ",,,,,,OAuth CID: 378", respectively. To improve our results, the under-sampling techniques consider the number of duplicated contents and the specified processing for the weighting.

### 4.2 Similar Contents between Positive and Negative Examples

To investigate in more detail, the number of false positive and false negative decreases from 314,835 to 62,381 and 1,582 to 382, respectively, after removing duplicated vectors from the matrices. We apply Multidimensional scaling (MDS), one of the techniques creating a map displaying the relative positions of a number of objects, to matrices combined vectors of false negative and false positive. Figure 4 depicts the result of MDS applied to the feature matrix of incorrect predicted data and shows there is no correlation between the false negative and false positive examples.

The both of the above analysis show features not derived from comments and user information, for example revision session ID used in [3], play an important role.

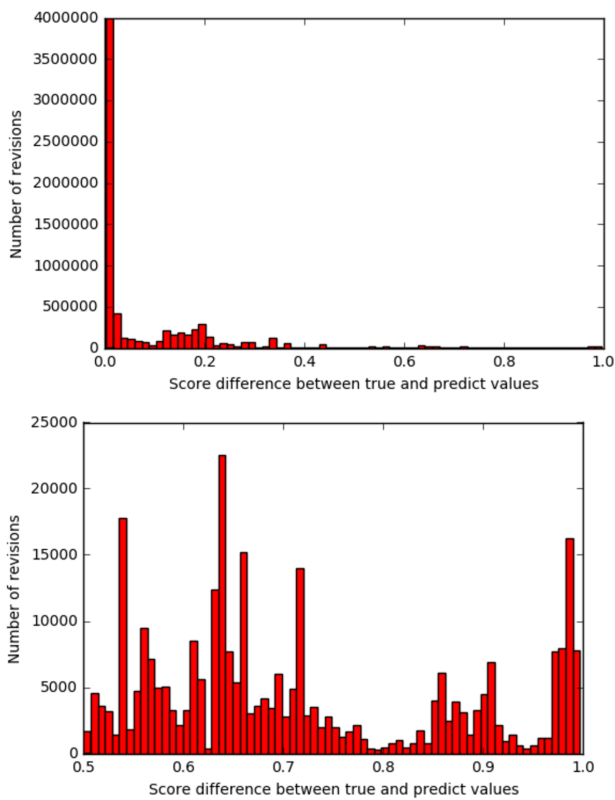

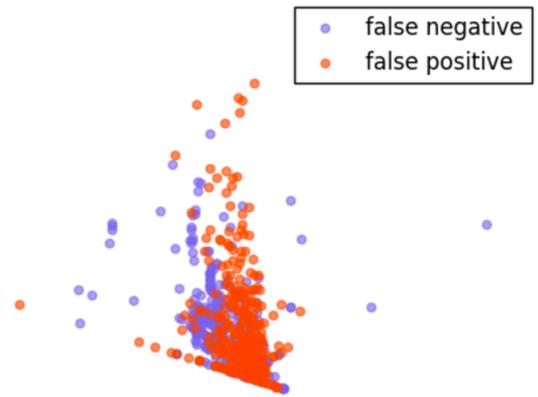

Figure 4: MDS of the feature matrix combined vectors of false negative and false positive.

Figure 3: The distribution of score difference between true values and predicted scores. The x-axis shows the difference between true values and predicted scores, and the y-axis shows the number of revisions. The upper and the lower figures depict entire results and results whose differences are greater than 0.5.

## 5. CONCLUDING REMARKS

In this paper, we introduce our method for the vandalism detection task of the WSDM Cup 2017 and confirm the validity by comparing between the result using some machine learning technique, for example under-sampling or stacking, and not using such technique.

Our final model results in the AUC-ROC of 0.95920 for validation data sets and unofficially 0.94412 for test data sets of WSDM Cup 2017.

We tuned hyper parameters of some final and stacking models with limited search grid spaces and we could not tuned them of all models we used, thus optimizing hyper parameters is in our future work. Moreover, new features have an impact of the score, thus it is also in our future work to create other features.